\documentclass{aa}
\usepackage{graphicx}
\usepackage{txfonts}
\usepackage{times,psfig}

\def\ltsima{$\; \buildrel < \over \sim \;$}
\def\lsim{\lower.5ex\hbox{\ltsima}}
\def\gtsima{$\; \buildrel > \over \sim \;$}
\def\gsim{\lower.5ex\hbox{\gtsima}}

\newcommand{\be}{\begin{equation}}
\newcommand{\en}{\end{equation}}

\begin{document}

\title{An effective selection criterion for GRB at high redshift with Swift}  
\title{Near real-time selection of high redshift GRBs with Swift}


\author{S.~Campana\inst{1}, G. Tagliaferri\inst{1}, D. Malesani\inst{1},
L. Stella\inst{1}, P. D'Avanzo\inst{1,4}, G. Chincarini\inst{1,5},
S. Covino\inst{1}} 

\authorrunning{S. Campana}

\titlerunning{High redshift GRBs selection}

\offprints{Sergio Campana, campana@merate.mi.astro.it}

\institute{INAF-Osservatorio Astronomico di Brera, Via Bianchi 46, I--23807
Merate (Lc), Italy
\and
International School for Advanced Studies (SISSA/ISAS), via Beirut 2-4, 34016
Trieste, Italy 
\and
INAF-Osservatorio Astronomico di Roma,
Via Frascati 33, I--00040 Monteporzio Catone (Roma), Italy
\and
Universit\`a degli Studi dell'Insubria, Dipartimento di Fisica e Matematica, Via Valleggio
11, I--22100 Como, Italy
\and
University of Milano--Bicocca, Department of Physics, Piazza delle Scienze 3,
20126 Milano, Italy 
}

\date{received, accepted}

\abstract{
A good fraction of GRBs detected by Swift are at a large redshift (up
to $z=6.3$, so far). Their study allows us to investigate, among other things,
the cosmic star formation in the early Universe (possibly up to the
re-ionization era) and the chemical enrichment of the high-redshift gas. Here
we present and discuss a method of selection which identifies high-redshift
candidates based only upon promptly-available information provided by
Swift. This method relies upon, Galactic extinction, GRB duration time and
absence of an optical counterpart in the UVOT telescope onboard Swift. This 
tool may provide an extremely effective way to locate high-redshift
astrophysical objects and to follow them in the optical/NIR band in near real
time. 
}

\maketitle

\section{Introduction}

The study of GRBs at high redshift holds great promises, allowing for the
investigation of GRB environments and their use to explore the galaxy 
formation and the chemical evolution up to the very young Universe. 

In almost two years of operations Swift has revolutionized the
fast-response multi-wavelength studies of gamma--ray bursts (GRBs). 
On average, the Burst Alert Telescope (BAT, 15--350 keV, Barthelmy et
al. 2005) localizes 2-3 GRBs per week ($2-3'$ precision at gamma--ray
energies); the X--Ray Telescope (XRT, 0.3--10 keV, Burrows et al. 2005)
localizes $\sim 70\%$  of them in $< 200$~s down to an accuracy of
$3-5\arcsec$ (Moretti et al. 2006). At the 
same time the Ultraviolet/Optical Telescope (UVOT, 170--600 nm, Roming et 
al. 2005) looks for optical counterparts. The GRB sample detected
by Swift has, on average, significantly dimmer optical afterglows than the GRB
sample of BeppoSAX and HETE II. In particular, about 1/3 have no optical
counterpart, despite deep and prompt searches. Indeed, some of these ``dark
bursts'' have relatively bright near-infrared (NIR) counterparts, 
implying that their afterglows are not just intrinsically faint, but
suffer also substantial obscuration (either from dust or neutral hydrogen) or
are at high redshift. The mean redshift of the Swift bursts is $\langle z 
\rangle = 2.8$, significantly larger than that measured for pre-Swift
events (Jakobsson et al. 2006). Population synthesis models suggest that
$\gsim 7-10\%$ of the \textit{detected} GRBs are at $z > 5$ (Bromm \& Loeb 2006;
Jakobsson et al. 2006). Given the possible preference of GRBs for
low-metallicity environments, the fraction at high redshift may be even higher
(Natarajan et al. 2005).

These findings support the possibility that a significant fraction of the
optically dark GRBs detected by Swift are at very high redshift. 
GRB050904 at $z = 6.29$ (Kawai et al. 2006; Totani et al. 2006;
Tagliaferri et al. 2005) might therefore be only the first discovered GRB of a
population of very high-redshift events. There are already five GRBs at $z >
5$. There might well be GRBs at comparable redshift or even higher redshift in
the present Swift sample, the optical/NIR afterglow of which could not be
detected due to observational difficulties involved in the identification of
the afterglows and in the acquisition of their spectra. For example, a
possible very high redshift event was GRB060116, for which we derived a
best-fit photometric redshift of $z_{\rm phot} = 6.7$, with a less-probable
solution at $z_{\rm phot} = 4.1$ (Grazian et al. 2006). An attempt to
derive a spectrum was unsuccessful, being the source too weak due to the large
delay (2.7 d) between the GRB and the observation (Fern\'andez-Soto et
al. 2006, in preparation).  This testifies for the importance of a fast
reaction.

Here we describe a simple and fast selection procedure able to spot
high-redshift candidate GRBs. This is based on a combination of the GRB
duration and on the lack of an optical counterpart in the early UVOT data in
low extinction locations in our Galaxy.  

\section{Selection procedure}

\begin{figure}	
\includegraphics[width=6.4cm,angle=-90]{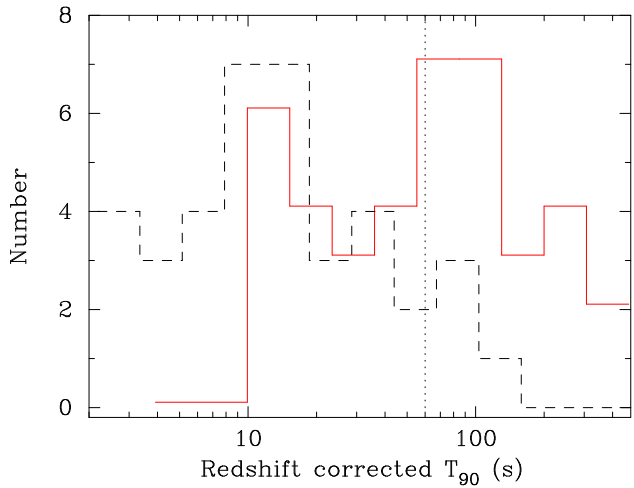}  
\caption{Intrinsic distribution of $T_{90}(z)$ durations observed by
Swift-BAT corrected for the cosmological time dilation (redshift). The
dashed-line histogram shows the observed distribution ($T_{90}(z)$), the
continuous-line histogram the same distribution shifted at $z=6$. The vertical
dotted line marks our reference constraint $T_{90}>60$ s.}
\label{fig1} 
\end{figure}

We start from the Swift sample of GRBs with known redshift (45 as of 30 Sep
2006). The distribution of redshift-corrected $T_{90}$ (i.e. the time
containing the $90\%$ of GRB fluence in the 15--350 keV BAT energy range) is shown
in Fig. 1. It is clearly apparent that long GRBs are relatively rare. However,
redshift increases the observed $T_{90}$. Taking the rest-frame distribution
of $T_{90}(z)$ at $z=6$, the fraction of observed GRBs with $T_{90}(z=6)>60$ s
would be $62\pm21\%$. By considering long GRBs with (observed) $T_{90}>60$ s
we can exclude a large fraction of GRB in the redshift interval 1--3 (see
Fig. 2), and selects $31\pm8\%$ of the GRB sample with known redshift and
$30\pm7\%$ of the full Swift sample (160 GRBs).
 
This is clearly not enough to identify high redshift GRBs. We add to this cut,
the lack of an UVOT counterpart, either in the first short (a few tens of
seconds) $V$ image. In order not to select GRB heavily absorbed in our Galaxy
we further request that the Galactic $E(B-V)<0.1$. UVOT observes shortward of
$5500\,\AA$ and it is virtually blind to objects at $z\gsim 5$ due to Lyman
drop out. The rationale behind this cut is to select highly absorbed or
high-redshift GRBs. The UVOT observing strategy described above (first short
$V$ image and then 100 s white image) has been implemented and used
systematically since March 2006. Since then 43 GRBs have been detected by
Swift. Of these 14 were discarded due to a too high Galactic extinction
($E(B-V)>0.1$). Of the 29 GRBs left, 17 do have an optical counterpart  
detected by UVOT ($59\pm25\%$).

\begin{figure}	
\includegraphics[width=6.4cm,angle=-90]{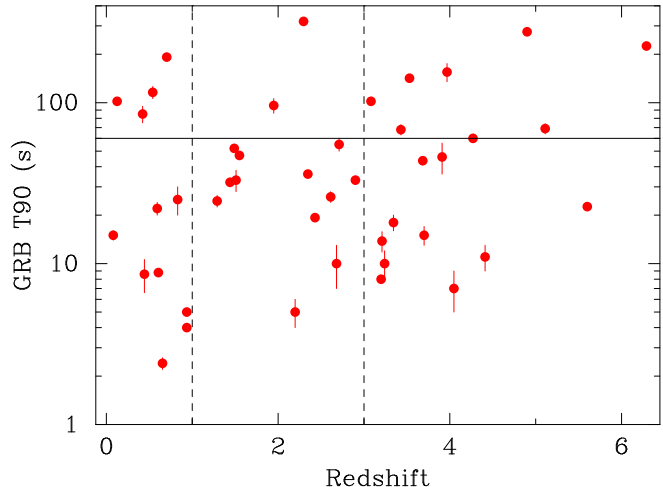}  
\caption{Redshift distribution of the complete sample of 41 Swift GRBs
versus their $T_{90}$, i.e. the time (redshift uncorrected) 
containing $90\%$ of the BAT 15--350 keV fluence. The
horizontal line marks our selection criteria for long-distant 
GRBs, and the two vertical dashed lines indicate the selected 
redshift interval. Basically, requiring $T_{90}>60$ s we mainly select 
closeby GRBs with long duration (like GRB060218 or GRB060614) or distant
($z>3$) GRBs.} 
\label{fig2} 
\end{figure}

\begin{figure}	
\includegraphics[width=6.4cm,angle=-90]{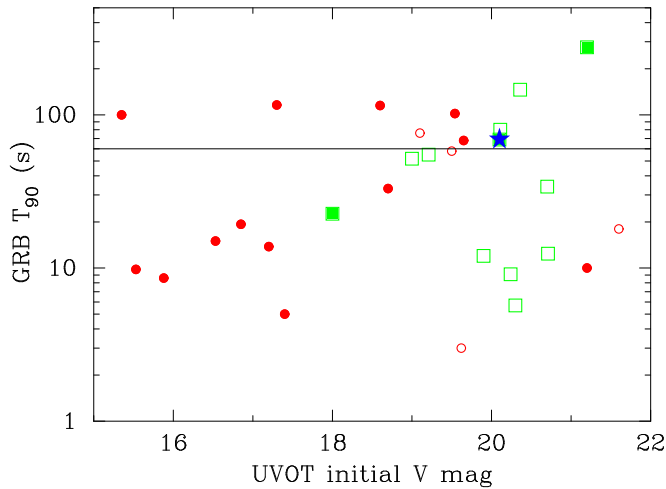} 
\caption{Distribution of UVOT $V$ magnitudes of all Swift GRBs from March 2006
versus their $T_{90}$ duration. Circles mark GRBs with a detection with UVOT
(in the $V$ band). Filled circles are GRBs with 
a spectroscopic redshift. Squares mark instead GRBs without 
a UVOT detection. Filled squares are GRBs with measured redshift. The star
marks GRB060522 with a non-detection in the $V$ band but a detection in white
light with UVOT.
The combination of $T_{90}>60$ s and absence of UVOT counterpart can be used
to single out high-redshift GRBs. We have 4 objects in this space of parameters:
two of them are without a secure redshift (they have strong upper
limits in the $R$-band and one has a $K$ band detection) and for two of them a
spectroscopic redshift with $z\gsim5$ has been measured.}
\label{fig2} 
\end{figure}

The intersection of the two constraints is extremely powerful. In the sample
of 29 GRBs we have 12 GRBs without a prompt UVOT counterpart and only 4 with
$T_{90}>60$ s (see Fig. 3). These are GRB060904A, GRB060814, GRB060522 and
GRB060510B. The last two GRBs have a spectroscopic redshift of $z=5.11$ and
$z=4.9$, respectively (Cenko et al. 2006; Price 2006). Actually, GRB060522 was
detected by UVOT in the 100 white light image at $19.65\pm0.21$ but not in the
400 s $V$ band exposure starting about 140 s after burst trigger with an upper
limit of $V>20.1$ (Fox et al. 2006; Holland 2006). GRB060904A was not detected
3.9 hr aftert he burst down to $R>22.0$ (Cenko \& Rau 2006). In the NIR Subaru
observations did not detect the afterglow down to $J>21.0$, 5 hr after the
burst (Aoki, Tanaka \& Kawai 2006). 
The $K-$band afterglow of GRB060814 was detected at $K\sim 18$, 4.4 hr after
the burst (Levan et al. 2006). VLT observations taken 58 min after the event
showed a $R\sim 24$ object consistent with the afterglow as well as other
closeby sources. The night was not photometric and the day after the same
object was $R=23.61\pm0.08$. ISIS photometry did not detect this object as
variable so we conclude that it likely is the host galaxy. 
These deep non-detections (or high magnitudes) in the $R$ band and
observations in the NIR suggest the presence of intrinsically heavily
absorbed afterglows or high-redshift afterglows. 

\section{Conclusions}

We have presented a powerful method to single out high-redshift GRBs
based on Swift promptly available data: GRB time duration ($T_{90}>60$ s),
lack of an optical counterpart in the UVOT data and low Galactic extinction
($E(B-V)<0.1$). These three constraints effectively select high-redshift GRBs. 
Basically, the first constraint preferentially selects distant GRBs whose
duration is stretched by cosmological time dilation. The second constraint
picks up highly extincted or high-redshift objects (UVOT observes shortward of
$5500\, \AA$, becoming blind at $z\gsim 5$ due to Lyman drop out). A further
constrain on the Galactic extinction cleans the sample from heavily absorbed
GRBs from our Galaxy.   
In a sample of 29 GRBs we are able to isolate four events. Of these two are at
a known redshift of $\sim 5$ (GRB060522 at $z=5.1$ and 060510B at $z=4.9$). Of
the remaining two, one has a large color index $R-K\gsim 5.6$ a few hours after 
the event and the other has firm upper limits in $R>22$ and in $J>21$ (about 5
hr after the burst). These observations indicate that three out of four
GRBs might be  high-redshift objects. GRB060814 having a relatively bright
host galaxy should instead be a highly reddened object. Our selection
procedure is effective in selecting confirmed high-redshift GRBs ($\gsim
50\%$) and at variance with other techniques is characterized by a low
contamination.  
This is very important because it will allow us to immediately identify the
few objects that deserve the fastest spectroscopy with an 8-m class telescope.

We stress that our selection criteria can be applied based on prompt
information provided by Swift only within a few hours (being the computation
of the $T_{90}$ contained in the refined BAT circular the main limitation). 
Additional criteria might be an X--ray column density consistent with the
Galactic value since the instrinsic column density is reduced roughly by
$\sim (1+z)^{2.6}$ (Campana et al. 2006; even if for GRB050904 we find a
larger column density with respect to the Galactic value, see Cusumano et
al. 2006). A further criterion is a relatively smooth prompt (BAT) light curve,
since at high redshift GRB spikes are dilated by cosmologial redshift.

High-redshift GRBs with a prompt follow up might yield a wealth of
unprecedented information on the chemical environment of the high-redshift
star-forming. In principle the metallicities (through Fe and Si column
densities), dust depletion (Zn abundances), the gas density and location with
respect to the GRB (through observation of fine-structure excited states) can
be traced. In addition, the star-formation history of the Universe up to the
re-ionization era ($z>6$) can be probed. Our technique may provide one of the
most effective ways to single out and study objects in the high-redshift Universe.

\begin{acknowledgements}
This work is supported at OAB--INAF by ASI grant I/R/039/04. 
\end{acknowledgements}

\end{document}